\documentclass[prl,reprint,superscriptaddress,showpacs]{revtex4-1}

\usepackage{graphicx}
\usepackage{amssymb,amsmath}
\usepackage{hyperref}
\usepackage{bigstrut}

\newcommand{\polpipn}{$\vec\gamma \vec p\to\pi^+n$ {}}

\begin{document}

\title{\boldmath First Measurement of the Polarization Observable $E$ in the
  $\vec p(\vec \gamma,\pi^+)n$ Reaction up to 2.25~GeV}


\newcommand*{\ANL}{Argonne National Laboratory, Argonne, Illinois 60439}
\newcommand*{\ANLindex}{1}
\affiliation{\ANL}
\newcommand*{\ASU}{Arizona State University, Tempe, Arizona 85287-1504}
\newcommand*{\ASUindex}{2}
\affiliation{\ASU}
\newcommand*{\BONN}{Universit\"at Bonn, 53115 Bonn, Germany}
\newcommand*{\BONNindex}{26b}
\affiliation{\BONN}
\newcommand*{\CSUDH}{California State University, Dominguez Hills, Carson, CA 90747}
\newcommand*{\CSUDHindex}{3}
\affiliation{\CSUDH}
\newcommand*{\CANISIUS}{Canisius College, Buffalo, NY}
\newcommand*{\CANISIUSindex}{4}
\affiliation{\CANISIUS}
\newcommand*{\CMU}{Carnegie Mellon University, Pittsburgh, Pennsylvania 15213}
\newcommand*{\CMUindex}{5}
\affiliation{\CMU}
\newcommand*{\CUA}{Catholic University of America, Washington, D.C. 20064}
\newcommand*{\CUAindex}{6}
\affiliation{\CUA}
\newcommand*{\SACLAY}{CEA, Centre de Saclay, Irfu/Service de Physique Nucl\'eaire, 91191 Gif-sur-Yvette, France}
\newcommand*{\SACLAYindex}{7}
\affiliation{\SACLAY}
\newcommand*{\UCONN}{University of Connecticut, Storrs, Connecticut 06269}
\newcommand*{\UCONNindex}{8}
\affiliation{\UCONN}
\newcommand*{\FU}{Fairfield University, Fairfield, CT 06824}
\newcommand*{\FUindex}{9}
\affiliation{\FU}
\newcommand*{\FIU}{Florida International University, Miami, Florida 33199}
\newcommand*{\FIUindex}{10}
\affiliation{\FIU}
\newcommand*{\FSU}{Florida State University, Tallahassee, Florida 32306}
\newcommand*{\FSUindex}{11}
\affiliation{\FSU}
\newcommand*{\Genova}{Universit$\grave{a}$ di Genova, 16146 Genova, Italy}
\newcommand*{\Genovaindex}{12}
\affiliation{\Genova}
\newcommand*{\GWUI}{The George Washington University, Washington, DC 20052}
\newcommand*{\GWUIindex}{13}
\affiliation{\GWUI}
\newcommand*{\ISU}{Idaho State University, Pocatello, Idaho 83209}
\newcommand*{\ISUindex}{14}
\affiliation{\ISU}
\newcommand*{\INFNFE}{INFN, Sezione di Ferrara, 44100 Ferrara, Italy}
\newcommand*{\INFNFEindex}{15}
\affiliation{\INFNFE}
\newcommand*{\INFNFR}{INFN, Laboratori Nazionali di Frascati, 00044 Frascati, Italy}
\newcommand*{\INFNFRindex}{16}
\affiliation{\INFNFR}
\newcommand*{\INFNGE}{INFN, Sezione di Genova, 16146 Genova, Italy}
\newcommand*{\INFNGEindex}{17}
\affiliation{\INFNGE}
\newcommand*{\INFNRO}{INFN, Sezione di Roma Tor Vergata, 00133 Rome, Italy}
\newcommand*{\INFNROindex}{18}
\affiliation{\INFNRO}
\newcommand*{\INFNTUR}{INFN, Sezione di Torino, 10125 Torino, Italy}
\newcommand*{\INFNTURindex}{19}
\affiliation{\INFNTUR}
\newcommand*{\ORSAY}{Institut de Physique Nucl\'eaire, CNRS/IN2P3 and Universit\'e Paris Sud, Orsay, France}
\newcommand*{\ORSAYindex}{20}
\affiliation{\ORSAY}
\newcommand*{\ITEP}{Institute of Theoretical and Experimental Physics, Moscow, 117259, Russia}
\newcommand*{\ITEPindex}{21}
\affiliation{\ITEP}
\newcommand*{\JMU}{James Madison University, Harrisonburg, Virginia 22807}
\newcommand*{\JMUindex}{22}
\affiliation{\JMU}
\newcommand*{\KNU}{Kyungpook National University, Daegu 702-701, Republic of Korea}
\newcommand*{\KNUindex}{23}
\affiliation{\KNU}
\newcommand*{\MCOLLEGE}{Montgomery College, Rockville, Maryland 20850}
\affiliation{\MCOLLEGE}
\newcommand*{\NRC}{NRC ``Kurchatov Institute'', PNPI, 188300, Gatchina,  Russia}
\affiliation{\NRC}
\newcommand*{\UNH}{University of New Hampshire, Durham, New Hampshire 03824-3568}
\newcommand*{\UNHindex}{24}
\affiliation{\UNH}
\newcommand*{\NSU}{Norfolk State University, Norfolk, Virginia 23504}
\newcommand*{\NSUindex}{25}
\affiliation{\NSU}
\newcommand*{\OHIOU}{Ohio University, Athens, Ohio  45701}
\newcommand*{\OHIOUindex}{26}
\affiliation{\OHIOU}
\newcommand*{\ODU}{Old Dominion University, Norfolk, Virginia 23529}
\newcommand*{\ODUindex}{27}
\affiliation{\ODU}
\newcommand*{\RPI}{Rensselaer Polytechnic Institute, Troy, New York 12180-3590}
\newcommand*{\RPIindex}{28}
\affiliation{\RPI}
\newcommand*{\ROMAII}{Universita' di Roma Tor Vergata, 00133 Rome Italy}
\newcommand*{\ROMAIIindex}{28}
\affiliation{\ROMAII}
\newcommand*{\MSU}{Skobeltsyn Institute of Nuclear Physics, Lomonosov Moscow State University, 119234 Moscow, Russia}
\newcommand*{\MSUindex}{29}
\affiliation{\MSU}
\newcommand*{\SCAROLINA}{University of South Carolina, Columbia, South Carolina 29208}
\newcommand*{\SCAROLINAindex}{30}
\affiliation{\SCAROLINA}
\newcommand*{\TEMPLE}{Temple University,  Philadelphia, PA 19122 }
\newcommand*{\TEMPLEindex}{31}
\affiliation{\TEMPLE}
\newcommand*{\JLAB}{Thomas Jefferson National Accelerator Facility, Newport News, Virginia 23606}
\newcommand*{\JLABindex}{32}
\affiliation{\JLAB}
\newcommand*{\UTFSM}{Universidad T\'{e}cnica Federico Santa Mar\'{i}a, Casilla 110-V Valpara\'{i}so, Chile}
\newcommand*{\UTFSMindex}{33}
\affiliation{\UTFSM}
\newcommand*{\EDINBURGH}{Edinburgh University, Edinburgh EH9 3JZ, United Kingdom}
\newcommand*{\EDINBURGHindex}{34}
\affiliation{\EDINBURGH}
\newcommand*{\GLASGOW}{University of Glasgow, Glasgow G12 8QQ, United Kingdom}
\newcommand*{\GLASGOWindex}{35}
\affiliation{\GLASGOW}
\newcommand*{\VT}{Virginia Tech, Blacksburg, Virginia 24061-0435}
\newcommand*{\VTindex}{36}
\affiliation{\VT}
\newcommand*{\VIRGINIA}{University of Virginia, Charlottesville, Virginia 22901}
\newcommand*{\VIRGINIAindex}{37}
\affiliation{\VIRGINIA}
\newcommand*{\WM}{College of William and Mary, Williamsburg, Virginia 23187-8795}
\newcommand*{\WMindex}{38}
\affiliation{\WM}
\newcommand*{\YEREVAN}{Yerevan Physics Institute, 375036 Yerevan, Armenia}
\newcommand*{\YEREVANindex}{39}
\affiliation{\YEREVAN}

\newcommand*{\NOWUK}{University of Kentucky, Lexington, Kentucky 40506}
\newcommand*{\NOWODU}{Old Dominion University, Norfolk, Virginia 23529}
\newcommand*{\NOWINFNGE}{INFN, Sezione di Genova, 16146 Genova, Italy}

\author {S.~Strauch}\affiliation{\SCAROLINA}
\author {W.J.~Briscoe} \affiliation{\GWUI}
\author{M.~D\"oring} \affiliation{\GWUI}
\author{E.~Klempt} \affiliation{\BONN}
\author{V.A.~Nikonov} \affiliation{\BONN} \affiliation{\NRC}
\author{E.~Pasyuk} \affiliation{\JLAB}
\author{D.~R\"onchen} \affiliation{\BONN}
\author{A.V.~Sarantsev} \affiliation{\BONN} \affiliation{\NRC}
\author{I.~Strakovsky} \affiliation{\GWUI}
\author{R.~Workman} \affiliation{\GWUI}

% ...

\collaboration{The CLAS Collaboration}
\noaffiliation

\date{\today}

\begin{abstract}
  First results from the longitudinally polarized frozen-spin target
  (FROST) program are reported.
  The double-polarization observable $E$, for the reaction $\vec
  \gamma \vec p \to \pi^+n$, has been measured using a circularly
  polarized tagged-photon beam, with energies from 0.35 to 2.37~GeV. 
  The final-state pions were detected with the CEBAF Large
  Acceptance Spectrometer in Hall B at the Thomas Jefferson National
  Accelerator Facility.  These polarization data agree fairly well
  with previous partial-wave analyses at low photon energies. Over
  much of the covered energy range, however, significant deviations
  are observed, particularly in the high-energy region where high-$L$
  multipoles contribute. The data have been included in new multipole
  analyses resulting in updated nucleon resonance parameters. We 
  report updated fits from the Bonn-Gatchina, J\"ulich, and SAID groups.
\end{abstract}


\maketitle


The spectrum of baryon resonances strongly depends on the internal
dynamics of its underlying constituents.  Recent lattice calculations
and quark models reveal a rich spectrum, in contrast to
phenomenological analyses of experiments, which have found a smaller
number of states \cite{Crede:2013kia,Klempt:2009pi}. The so-called
missing resonances have stimulated alternative interpretations of the
resonance spectrum. These include the formation of quasi-stable
diquarks \cite{Anselmino:1992vg}, string models running under the
acronym AdS/QCD \cite{Brodsky:2006uq}, models assuming some baryon
resonances are dynamically generated from the unitarized interaction
among ground-state baryons and mesons \cite{Kolomeitsev:2003kt}, and the speculation that a
phase transition may occur in high-mass excitations
\cite{Afonin:2007mj}. The photoproduction of mesons off nucleons
provides an opportunity to distinguish among these alternatives.

Four complex amplitudes govern the photoproduction of single pions, and a 
{\it complete} experiment requires the measurement of at least eight well-chosen
observables at each energy and production angle for both isospin-related 
reactions $\gamma p\to \pi^0p$  and  $\gamma p\to \pi^+n$.  However,
the current database for pion photoproduction is populated mainly by 
unpolarized cross sections and single-spin observables, which do not form
a complete experiment. This is 
particularly true for $\pi^+n$ photoproduction at photon energies
above 1.8~GeV. This incompleteness 
of the database leads to ambiguities in the multipole solutions.

In this Letter we present a measurement of the double-polarization
observable $E$ in the \polpipn reaction of circularly polarized
photons with longitudinally polarized protons. The polarized cross
section is in this case given by \cite{Barker:1975bp}
\begin{equation}
  \left(\frac{d\sigma}{d\Omega}\right) = 
  \left(\frac{d\sigma}{d\Omega}\right)_{0} 
  \left(1 - P_z P_{\odot} E \right),
\end{equation}
where $(d\sigma/d\Omega)_0$ is the unpolarized cross section; $P_z$
and $P_\odot$ are the target and beam polarizations, respectively.
The observable $E$ is the helicity asymmetry of the cross section,
\begin{equation}
  E = \frac{d\sigma_{1/2} - d\sigma_{3/2}}{d\sigma_{1/2} 
    + d\sigma_{3/2}}
\end{equation}
for aligned, total helicity $h = 3/2$, and anti-aligned, $h = 1/2$,
photon and proton spins.  These data are fitted using 
three very different PWA models from the
Bonn-Gatchina~\cite{Anisovich:2011fc},
Bonn-J\"ulich~\cite{Ronchen:2014cna}, and GWU~\cite{Workman:2012jf} groups. 
The resulting consistency of helicity amplitudes for the
dominant resonances demonstrates that the PWA results are largely driven by the data alone; the
modest differences gauge the model-dependence. This consistency provides 
an excellent starting point to search for new resonances. 

Earlier measurements have been reported for the polarization
observable $E$ in the $\pi^0p$ channel \cite{Gottschall:2013uha} and
some cross-section helicity-asymmetry data exists in the $\pi^+n$
channel \cite{Ahrens:2006gp,Ahrens:2004pf}. Here we report $E$
measurements of unprecedented precision covering, for the first time,
nearly the entire resonance region.

The experiment was performed at the Thomas Jefferson National
Accelerator Facility (JLab).  Longitudinally polarized electrons from
the CEBAF accelerator with energies of $E_e = 1.645$~GeV and 2.478~GeV
were incident on the thin radiator of the Hall-B Photon Tagger
\cite{Sober:2000we} and produced circularly polarized tagged photons
in the energy range between $E_\gamma = 0.35$~GeV and 2.37~GeV. 

The degree of circular polarization of the photon beam, $P_\odot$,
depends on the ratio $x=E_{\gamma}/E_e$ and increases from zero to the
degree of incident electron-beam polarization, $P_e$, monotonically
with photon energy \cite{Olsen:1959zz}
\begin{equation}
  P_\odot = P_e \cdot \frac{4x - x^2}{4 - 4x + 3x^2}.
\end{equation}
Measurements of the electron-beam polarization were made routinely
with the Hall-B M{\o}ller polarimeter.  The average value of the
electron-beam polarization was found to be $P_e = 0.84\pm 0.04$.  The
electron-beam helicity was pseudo-randomly flipped between $+1$ and
$-1$ with a 30~Hz flip rate.  

The collimated photon beam irradiated a frozen-spin target (FROST)
\cite{Keith201227} at the center of the CEBAF Large Acceptance
Spectrometer (CLAS) \cite{Mecking:2003zu}.  Frozen beads of butanol
(C$_4$H$_9$OH) inside a 50~mm long target cup were used as target
material.  The protons of the hydrogen atoms in this material were
dynamically polarized along the photon-beam direction.  The degree of
polarization was on average $P_z = 0.82\pm 0.05$.  The proton
polarization was routinely changed from being aligned along the beam
axis to being anti-aligned.  Quasi-free photoproduction off the
unpolarized, bound protons in the butanol target constituted a
background.  Data were taken simultaneously from an additional carbon
target downstream of the butanol target to allow for the determination
of this bound-nucleon background.  A small unpolarized hydrogen
contamination of the carbon target has been corrected for in the
analysis.

Final-state $\pi^+$ mesons were detected in CLAS.  The particle
detectors used in this experiment were a set of plastic scintillation
counters close to the target to measure event start times (start
counter) \cite{Sharabian:2005kq}, drift chambers \cite{Mestayer:2000we}
to determine charged-particle trajectories in the magnetic field
within CLAS, and scintillation counters for flight-time measurements
\cite{Smith:1999ii}.  Coincident signals from the photon tagger,
start-, and time-of-flight counters constituted the event trigger.
Data from this experiment were taken in seven groups of runs with
various electron-beam energies and beam/target polarization
orientations.  Events with one and only one positively charged
particle and zero negatively charged particles detected in CLAS were
considered.  The $\pi^+$ mesons were identified by their charge (from
the curvature of the particle track) and by using the time-of-flight
technique.  Photoproduced lepton-pair production in the nuclear
targets was a forward peaked background.  This background was strongly
suppressed with a fiducial cut on the polar angle of the pion,
$\theta^\text{lab}_\pi > 14^\circ$.  

\begin{figure}[htb!]
  \begin{center}
    \includegraphics[width=\linewidth]{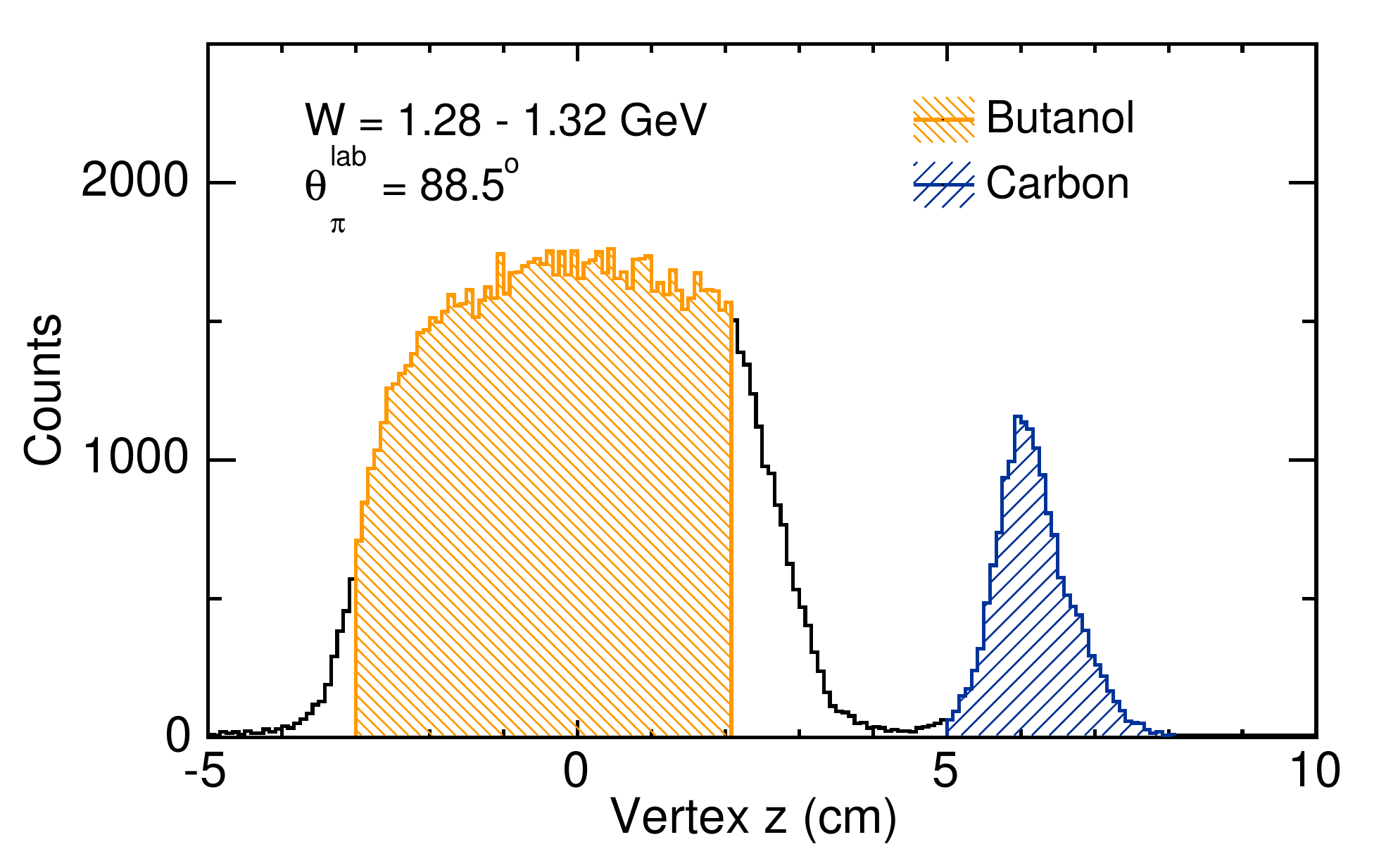}
  \end{center}
  \caption{Example of a reconstructed distribution of the reaction
    vertex along the beam line for
    events at $W \approx 1.30$~GeV and $\theta^{\rm lab} \approx
    88.5^\circ$ originating in the butanol and carbon
    targets.  The shaded areas indicate the $z$-vertex ranges used
    in the analysis. 
    \label{fig:mvrt_z}}
\end{figure}

\begin{figure}[htb!]
  \begin{center}
    \includegraphics[width=\linewidth]{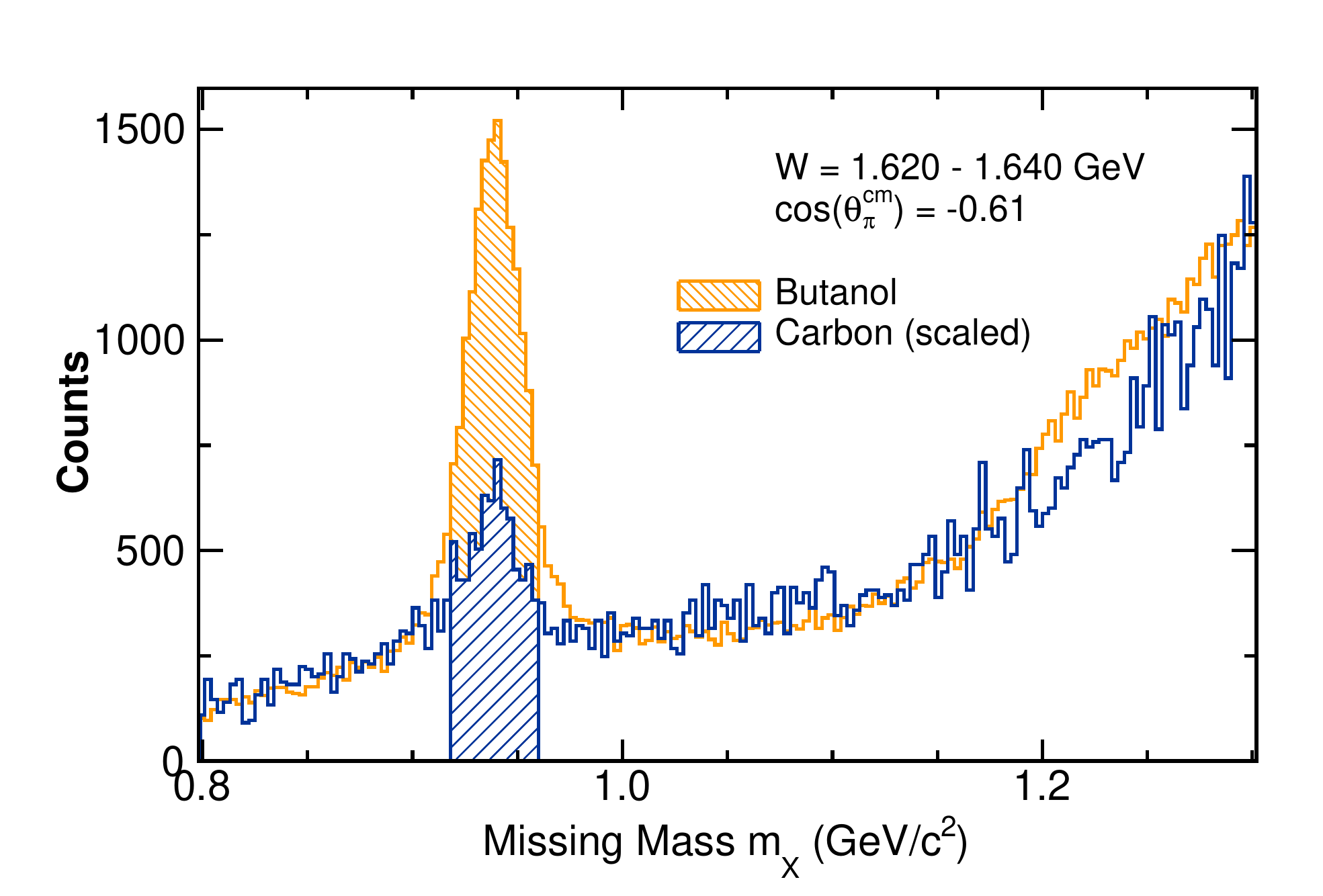}
    \includegraphics[width=\linewidth]{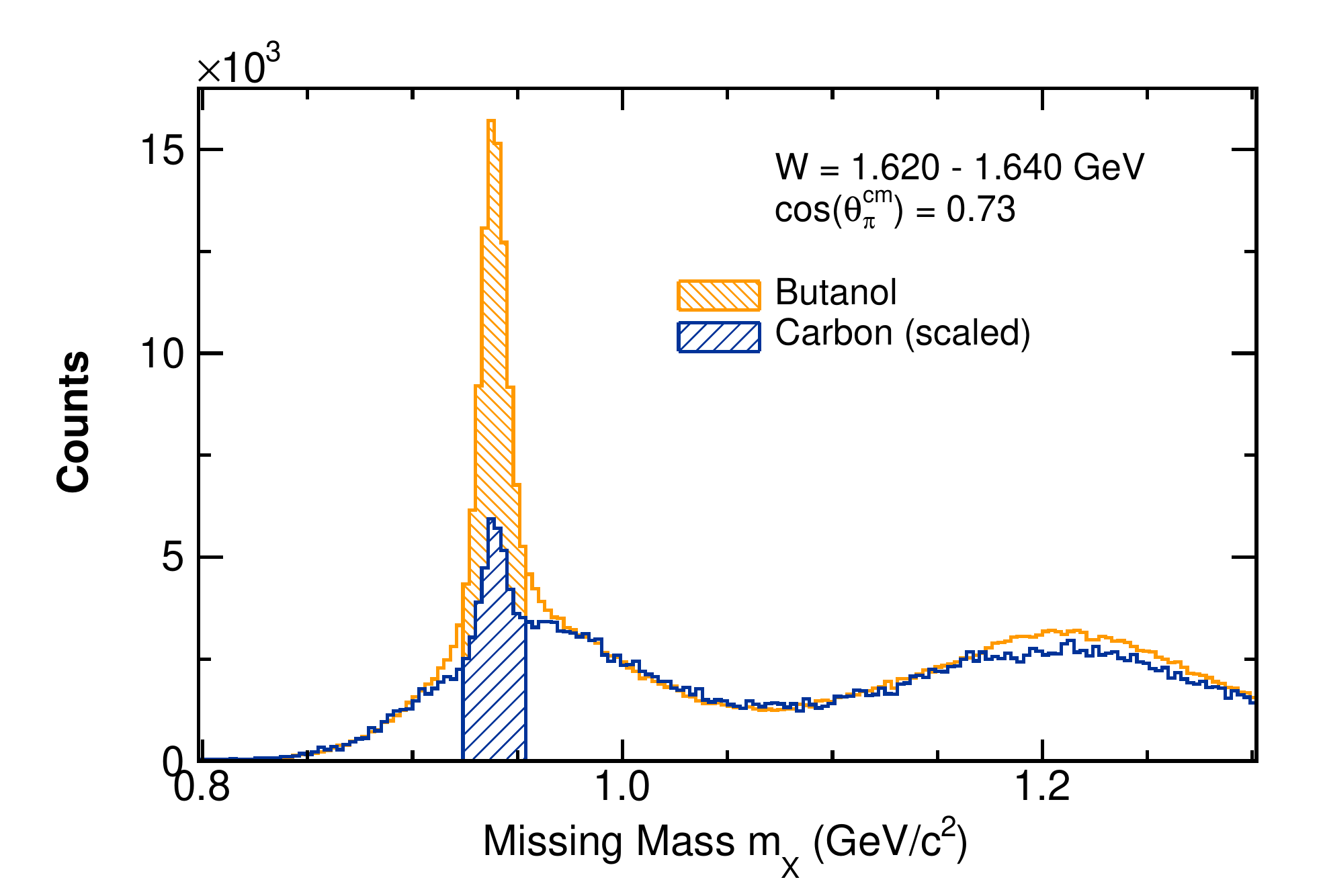}
  \end{center}
  \caption{(Color online) Examples of butanol missing-mass
    distributions, $\gamma p \to \pi^+X$, overlaid with scaled
    distributions from the carbon-target.  The hatched region selects
    the butanol- and carbon-target events which were used in the
    subsequent analysis.  The butanol yield at larger missing masses
    contains multi-pion final-state events off the free proton and
    exceed the carbon yield.}
  \label{fig:mmX}
\end{figure}

The observable $E$ was determined in 900 kinematic bins of $W$ and
$\cos \theta_\pi^\text{cm}$, where $W$ is the center-of-mass energy
and $\theta_\pi^\text{cm}$ is the pion center-of-mass angle with
respect to the incident photon momentum direction.  For each bin three
missing-mass distributions in the $\gamma p \to \pi^+ X$ reaction were
accumulated: for events originating in the butanol-target with a total
helicity of photons and polarized protons of $h = 3/2$, for butanol
events with $h = 1/2$, and for events originating in the
carbon-target. The production target
was identified by the reconstructed position of the reaction vertex;
see~Fig.~\ref{fig:mvrt_z}.  
To determine the bound-nucleon background in the
butanol data, the carbon-data distribution was scaled by a factor
$\alpha$ to fit the butanol
missing-mass distribution up to 1.05~GeV/$c^2$, together with a
Gaussian peak. Over all kinematic bins, the average value of $\alpha$
is 5. Examples of two
angular bins at $W \approx 1.63$~GeV are shown in Fig.~\ref{fig:mmX}.
The number of events, $N_{3/2}^B$, $N_{1/2}^B$, and $N^C$, for a given
kinematic bin were then selected by the condition $|m_X - m_0| <
2\sigma_H$, where $m_0$ and $\sigma_H$ are the peak position and peak
width of the neutron in the missing mass distribution taken from
the fit.  The selection is indicated by the hatched region in
Fig.~\ref{fig:mmX}.

The observable $E$ was finally extracted from the polarized yields, $N^p_{3/2}$ and
$N^p_{1/2}$, of \polpipn events for total helicities $h=3/2$ and
$1/2$, respectively, and the average beam and target
polarizations, 
\begin{equation}
  E = \frac{1}{\overline P_z \overline P_\odot}\left(\frac{N^p_{1/2}
      -N^p_{3/2}}{N^p_{1/2}+N^p_{3/2}}\right).
  \label{eq:estimate}
\end{equation}
As the bound nucleons in the butanol target are unpolarized, the
helicity difference in the event numbers is due only to the polarized
hydrogen, $N^p_{1/2} - N^p_{3/2} = N^B_{1/2} - N^B_{3/2}$.  The total
yield from polarized hydrogen was determined from the butanol and
carbon yields, $N^p_{1/2} + N^p_{3/2} = (N^B_{1/2} +N^B_{3/2} - \alpha
N^C)\kappa$, where $\kappa = 1.3$ is an experimentally well
determined correction factor which takes into account the hydrogen
contamination of the carbon target and the limited resolution in the
target reconstruction at very forward pion angles.  The experimental
value for $E$ is then given by
\begin{equation}
  \label{eq:E}
  E = \frac{1}{\overline P_z \overline P_\odot\kappa}\left[\frac{N^B_{1/2}
      -N^B_{3/2}}{N^B_{1/2}+N^B_{3/2}  - \alpha N^C}\right].
\end{equation}

The statistical uncertainty of $E$ is determined from the counting
statistics of the event yields and from the statistical uncertainty of the
scale factor $\alpha$.  The relative systematic uncertainty is
dominated by the uncertainty in the product of the beam and target
polarizations, about $\pm 7.5$\%.  The hydrogen contamination contributes
with $\pm 1.5$\%.  Point-to-point uncertainties are due to the background
subtraction, $\pm 0.03$, and, only at the most forward pion angles, due to
the limited vertex resolution, an additional contribution $< 0.015$.

\begin{figure*}[htb!]
  \begin{center}
    \includegraphics[width=\linewidth]{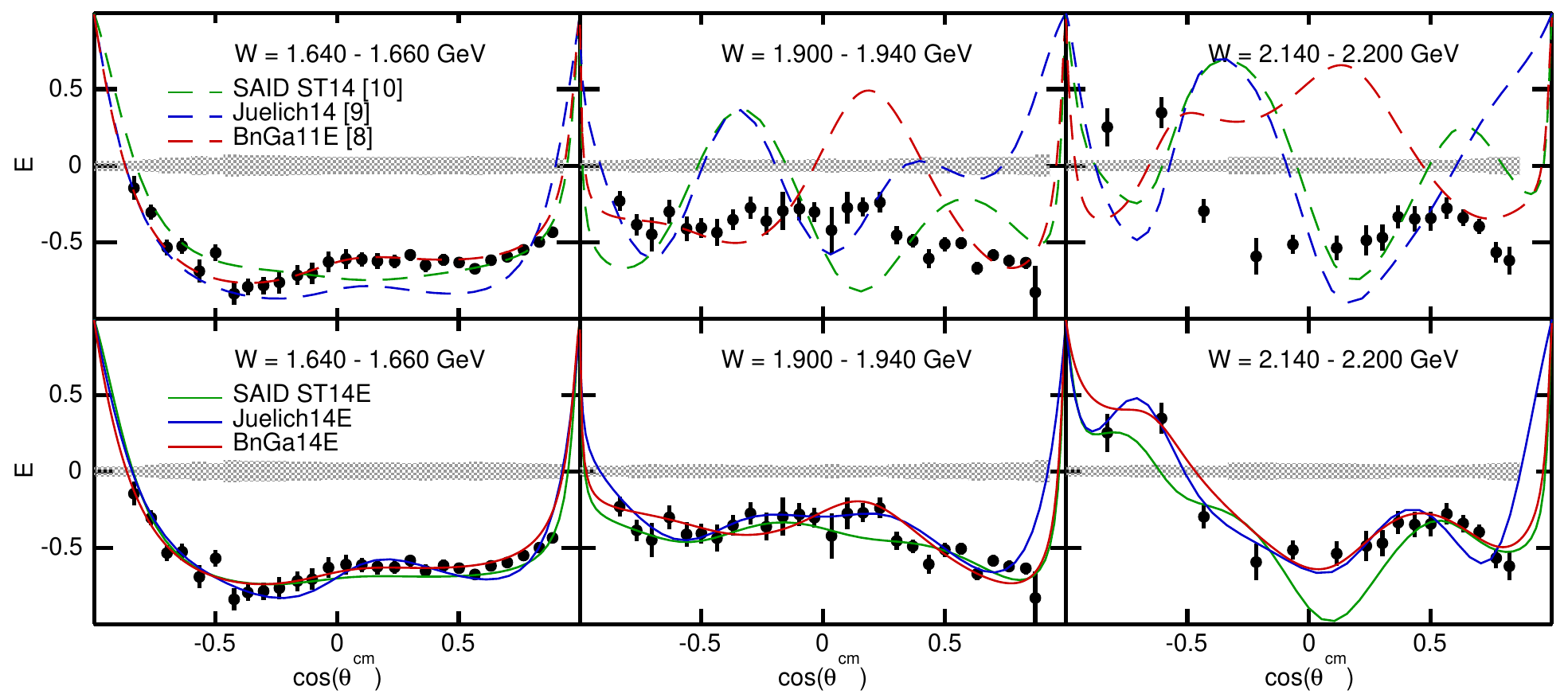}
  \end{center}
  \caption{(Color online) Double polarization observable $E$  in the
    \polpipn reaction as a function of
    $\cos\theta_\pi^\text{cm}$ for three selected bins of the center-of-mass
    energy $W$.  Systematic uncertainties are indicated as shaded
    bands.  The curves in the upper panels are results from the SAID
    ST14 \cite{Workman:2012jf}, J\"ulich14 \cite{Ronchen:2014cna}, and
    BnGa11E \cite{Anisovich:2011fc} analyses.  The curves in the lower
    panels are results from updated analyses including the present $E$ data.
    \label{fig:all}}
\end{figure*}

The angular distributions, plotted in Fig.~\ref{fig:all} as functions
of $\cos\theta_\pi^\text{cm}$, display an approximate 'U'-shaped
distribution between the required maxima at $\cos\theta_\pi^\text{cm}
= \pm 1$ and dipping to about $-0.5$ for energies up to about $W =
1.7$~GeV. This differs from the $E$ measurements for $\pi^0 p$
photoproduction from CBELSA-TAPS~\cite{Gottschall:2013uha}. There, in
a broad energy bin covering 960 -- 1100~MeV, one sees a zero crossing
near 90 degrees. In general, for the $\pi^+n$ final state and
$W<1.5$~GeV, the data are well
predicted~\cite{Anisovich:2011fc,Ronchen:2014cna,Workman:2012jf}, as
Fig.~\ref{fig:all} shows, because the analyses are constrained by
older MAMI-B data~\cite{Ahrens:2006gp}. However, at most of the higher
photon energies, where no similar constraints exist, the BnGa,
J\"ulich, and SAID analyses predict more pronounced angular variations
than are seen in the data.  These qualitative features exist in the
MAID~\cite{Drechsel:2007if} predictions as well.

Given the relative lack of polarization data at the highest
energies, it is not surprising that a much better fit to these 
new $E$ measurements is achieved once they
are included in the database. In principle, a fit may be achieved 
through small amplitude changes that 
produce large changes in the polarization observables, 
through a substantial modification of the assumed
resonance and background contributions, or through 
the addition of new resonances. Having the BnGa, J\"ulich and SAID
analyses together
we are able to compare results with a minimal set
of resonances (SAID) to the larger sets required in the
BnGa and J\"ulich analyses.

To show the impact of the new $E$ data, Table~\ref{tbl-1} shows the
helicity couplings of selected low-mass nucleon resonances before and
after including the data in the three analyses.  The baseline SAID and
J{\"u}lich fits were done with the same updated database to have a
common point of comparison.  The SAID and BnGa analyses compare
changes in the Breit-Wigner resonance photo-decay parameters, while the
J\"ulich results determine photo-couplings at the pole. While these
quantities are different in principle, a recent
study~\cite{Workman:2013rca} has found qualitative agreement between
the moduli of pole residues and real Breit-Wigner
quantities. Comparisons between the two sets will be made at this
qualitative level.

The SAID resonance couplings have changed only
slightly for most states, usually within the estimated uncertainties
of the extraction. As no new states are explicitly added, the fit
below the highest energies has been accomplished with only small
changes to the existing states. For the highest energies, unambiguous
resonance extraction is complicated by a number of factors. Here, the
non-resonant background is significant, as can be seen from the
dominant forward peaking in the cross section
\cite{Buschhorn:1967zz}. In addition, one must deal with the
interference of many amplitudes of a similar size, with resonances
tending to be coupled only weakly to the $\pi N$ channel.
\begin{table*}
  \centering
  \caption{Fits to the new CLAS data (labeled $E$) and predictions. Breit-Wigner helicity amplitudes 
    for the SAID (ST14 based on CM12~\cite{Workman:2012jf}) and
    Bonn-Gatchina (\cite{Gottschall:2013uha}; $^\dagger$:
    entries from Ref.~\cite{Anisovich:2011fc}) analyses.
    Values from J\"ulich (J\"ulich14 based on Ref.~\cite{Ronchen:2014cna}) are quoted 
    at the $T$-matrix pole including the complex phase in
    parentheses. Helicity amplitudes $A_{1/2}$ and $A_{3/2}$ are given in units of 
    (GeV)$^{-1/2} \times 10^{-3}$.}
  \begin{tabular}{lc|r@{$\pm$}l		
      r@{$\pm$}l		
      r@{$($}l		
      r@{$($}l		
      r@{$\pm$}l		
      r@{$\pm$}l}		
    \hline
    \hline
    &&\multicolumn{2}{c}{ST14} 
    & \multicolumn{2}{c}{ST14{\bf E}} 
    & \multicolumn{2}{c}{J\"ulich14} 
    & \multicolumn{2}{c}{J\"ulich14{\bf E}} 
    & \multicolumn{2}{c}{BnGa11E} 
    & \multicolumn{2}{c}{BnGa14{\bf E}} 
    \bigstrut\\
    \hline
    N(1440)1/2$^+$              & $A_{1/2}$     & $-65$&$5$     & $-60$&$5$     & $-56$&$\,\,\,+5^\circ)$& $-53$&$\,\,\,-6^\circ)$      & $-62$&$ 8$                    & $-60 $&$ 8$    \bigstrut[t]\\
    N(1520)3/2$^-$              & $A_{1/2}$     & $-22$&$2$     & $-24$&$2$     & $-25$&$-13^\circ)$    & $-22$&$-14^\circ)$    & $-20$&$ 3$                    & $-24 $&$ 4$    \\
    & $A_{3/2}$                                 & $142$&$5$     & $138$&$3$     & $112$&$+28^\circ)$    & $104$&$+22^\circ)$    & $131$&$ 7$                    & $130 $&$ 6$    \\
    N(1535)1/2$^-$              & $A_{1/2}$     & $115$&$10$    & $120$&$10$    & $ 52$&$-14^\circ)$    & $ 51$&$-20^\circ)$    & $105$&$ 9$                    & $100 $&$ 12$   \\
    N(1650)1/2$^-$              & $A_{1/2}$     & $ 55$&$30$    & $ 60$&$30$    & $ 28$&$\,\,\,+7^\circ)$& $ 30$&$-21^\circ)$   & $ 33$&$ 7$                    & $ 32 $&$ 6$    \\
    $\Delta$(1620)1/2$^-$       & $A_{1/2}$     & $ 35$&$5$     & $ 30$&$5$     & $ 23$&$+14^\circ)$    & $ 25$&$+13^\circ)$    & $ 52$&$ 5$                    & $ 59 $&$ 8$    \\ 
    $\Delta$(1700)3/2$^-$       & $A_{1/2}$     & $128$&$20$    & $150$&$20$    & $118$&$\,\,\,-6^\circ)$& $121$&$-14^\circ)$   & $160$&$ 20^\dagger$           & $165 $&$ 20$   \\
    & $A_{3/2}$                                 & $ 91$&$30$    & $110$&$30$    & $106$&$+20^\circ)$    & $116$&$+52^\circ)$    & $165$&$ 25^\dagger$           & $170 $&$ 25$   \\
    $\Delta$(1905)5/2$^+$       & $A_{1/2}$     & $ 30$&$6$     & $ 30$&$5$     & $ 13$&$+17^\circ)$    & $-39$&$+26^\circ)$    & $ 25$&$  5^\dagger$           & $ 30 $&$ 8$    \\
    & $A_{3/2}$                                 & $-70$&$10$    & $-50$&$10$    & $-79$&$-59^\circ)$    & $-49$&$-67^\circ)$    & $-49$&$  4^\dagger$           & $-50 $&$ 5$    \\ 
    $\Delta$(1950)7/2$^+$       & $A_{1/2}$     & $-70$&$5$     & $-80$&$5$     & $-70$&$-15^\circ)$    & $-64$&$-16^\circ)$    & $-70$&$ 5$                    & $-68 $&$ 5$    \\
    & $A_{3/2}$                                 & $-90$&$5$     & $-90$&$5$     & $-86$&$\,\,\,-8^\circ)$& $-91$&$\,\,\,-7^\circ)$      & $-93$&$ 5$                    & $-94 $&$ 4$    \bigstrut[b]\\
    \hline \hline
  \end{tabular}
  \label{tbl-1}
\end{table*}

The results given in Table~\ref{tbl-1} can be compared in detail with a similar
table presented in the CBELSA-TAPS collaboration analysis of $E$ data
for $\pi^0 p$ photoproduction~\cite{Gottschall:2013uha}.  Here the
BnGa11E column gives the result of including these new $\pi^0 p$ $E$
data in a fit. 
As the BnGa11E fit changed very little, these values
(indicated with daggers) have been taken from the BnGa2011
solution~\cite{Anisovich:2011fc}. Comparison with the fit ST14E is
interesting in that almost all helicity amplitudes agree with those
from BnGa11E, within quoted errors.

Including the new $E(\pi^0p)$ data~\cite{Gottschall:2013uha} in the 
J\"ulich14 analysis led to an improved prediction of the $E(\pi^+n)$ data at
intermediate energies but still failed to predict the new data at high
energies (cf.~Fig.~\ref{fig:all}).  The impact of the new $E(\pi^+n)$ data on some
resonance parameters is significant in the J\"ulich14E
re-analysis. For the N(1650)1/2$^-$ the phase changes by 28$^\circ$,
but also the SAID analysis finds that this helicity coupling is not
well determined.  The N(1535)1/2$^-$ helicity coupling is small
because that resonance is narrower than in other
analyses~\cite{Ronchen:2014cna}.  For some prominent resonances, such
as the Roper, the N(1520)3/2$^-$, the $\Delta$(1232)3/2$^+$, and the
$\Delta$(1950)7/2$^+$, the $E$ data change the modulus and complex
phase of the helicity couplings only moderately by around 10\%. In
contrast, for less prominent and more inelastic resonances, changes
can be much larger as in case of the $\Delta$(1905)5/2$^+$.
In the J\"ulich14E solution, changes in very high-$L$
multipoles are larger than for the SAID analysis. Through 
correlations, high multipoles induce changes in lower multipoles. This 
explains why the new data has a larger impact for the 
J\"ulich analysis than for the SAID analysis.

One poorly known state, the $\Delta (2200) {7 \over 2}^-$, emerges
and plays an important role in improving the Bonn-Gatchina fit at
the highest energies~\cite{BnGaResult}. This state also exists in the
Bonn-Juelich analysis, but is not included in the SAID analysis. If this
state exists, it would be in plain conflict with the prediction of models
assuming a phase transition in high-mass resonances.

In summary, we have presented measurements of the double-polarization
observable E in the \polpipn reaction up to $W$ = 2.3 GeV over a large
angular range.  These results are the first of the FROST program at
JLab.  The fine binning and unprecedented quantity of the data impose
tight constraints on partial-wave analysis, especially at high-$L$
multipoles and at high center-of-mass energies where new resonances
are expected to exist.  These more tightly constrained amplitudes help
to fix the $\pi N$ components of larger multi-channel analyses as
well.  The SAID and Bonn-Gatchina solutions found minor changes of
helicity couplings for most resonances, while the new $E$ data led to
major changes for the J\"ulich solution and indications for a new
state in the BnGa re-analysis.

\begin{acknowledgments}
  The authors gratefully acknowledge the work of the Jefferson Lab
  staff.  This work was supported by the U.S.  National Science
  Foundation, the U.S. Department of Energy (DOE), the Chilean
  Comisi\'on Nacional de Investigaci\'on Cient\'ifica y Tecnol\'ogica
  (CONICYT), the Deutsche Forschungsgemeinschaft (SFB/TR16), the
  French Centre National de la Recherche Scientifique and Commissariat
  \`{a} l'Energie Atomique, the Italian Istituto Nazionale di Fisica
  Nucleare, the National Research Foundation of Korea, the Russian
  Foundation of Fundamental Research, the Russian Science
  Foundation (RNF), and the UK Science and Technology Facilities
  Council (STFC).  Jefferson Science Associates, LLC, operates the
  Thomas Jefferson National Accelerator Facility for the United States
  Department of Energy under contract DE-AC05-060R23177.
\end{acknowledgments}
\bibliography{g9a_piplus_E}

\begin{thebibliography}{24}
\bibitem [{BnG()}]{BnGaResult}%
  \BibitemOpen
  \href@noop {} {}\bibinfo {note} {{Details of this coupled-channel analysis
  are presented in a follow up paper}}\BibitemShut {NoStop}%
\end{thebibliography}

\end{document}